\documentstyle[preprint, aps]{revtex}

\tightenlines

\begin{document}
\draft
\preprint{ND Atomic Theory Preprint 98/11}
\title{
High-precision calculations of dispersion coefficients,
static  dipole polarizabilities, and atom-wall interaction
constants for alkali-metal atoms
}

\author{A. Derevianko$^{1,2}$, 
W. R. Johnson$^{1,2}$, M. S. Safronova$^{1}$, and J. F. Babb$^{2}$}
\address{${}^1$Department of Physics, Notre Dame University,
Notre Dame, Indiana  46556\\
${}^2$Institute for Theoretical Atomic and Molecular Physics\\
Harvard-Smithsonian Center for Astrophysics,
Cambridge, Massachusetts 02138}

\date{\today}
\maketitle

\begin{abstract}
The van der Waals coefficients for the
alkali-metal atoms from Na to Fr
interacting in their ground states, are calculated
using relativistic {\em ab initio\/} methods.
The accuracy of the calculations is estimated  by also evaluating 
atomic static electric dipole polarizabilities and
coefficients for the interaction of the atoms
with a perfectly conducting wall.
The results are in excellent agreement
with the latest data from ultra-cold collisions
and from 
studies of magnetic field induced  Feshbach resonances in Na and Rb.
For Cs we provide critically needed data for ultra-cold 
collision studies.
\end{abstract}

\pacs{PACS: 34.20.Mq, 32.10.Dk, 34.50.Dy, 31.15.Ar}

The van der Waals interaction plays an important role in
characterizing ultra-cold  collisions between two
ground state alkali-metal atoms.  While the calculation of 
interaction coefficients  has been a subject of great interest
in atomic, molecular and chemical physics for a very long time, it is
only very recently that novel cold collision experiments,
photoassociation spectroscopy, and analyses of magnetic field induced
Feshbach resonances have yielded strict constraints on magnitudes of
the coefficients. Moreover, due to the extreme sensitivity of elastic
collisions to the long-range part of the potentials, knowledge of the
van der Waals coefficients influences predictions of signs and magnitudes
of scattering lengths.  Although many theoretical methods have been developed
over the years to calculate van der Waals coefficients, 
persistent discrepancies remain.

In this paper, various relativistic {\em ab initio} methods are
applied to determine the van der Waals coefficients for the
alkali-metal dimers of Na to Fr~\cite{Yan98}.  As a check on our
calculations, we also evaluate the atom-wall interaction constants, 
which have
recently been calculated by other methods, and use them as a sensitive
test of the quality of our wave functions. Furthermore, we  calculate
atomic polarizabilities and compare them to experimental data, where
available.

The dynamic polarizability at imaginary frequency $\alpha(i \omega)$ 
for a valence state $|v\rangle$
can
be represented as a sum over intermediate states $|k\rangle$

\begin{equation}
\alpha \left( i\omega \right) =\frac{2}{3}\sum_{k}\frac{E_{k}-E_{v}}{%
\left( E_{k}-E_{v}\right) ^{2}+\omega ^{2}}
\langle v |{\mathbf R}|k \rangle
\cdot \langle k  |{\mathbf R}|v \rangle \, ,
\label{EqPol}
\end{equation}
where 
the sum includes an integration over  continuum states and  
${\mathbf R} = \sum_{j=1}^N {\mathbf r}_j$ is the dipole operator for
the $N$-electron atomic system. We use
atomic units throughout.  
The dispersion coefficient $C_6$ of the van der Waals interaction
between two identical atoms is 
\begin{equation}
C_6 = \frac{3}{ \pi} \int_0^\infty d\omega \; [\alpha( i \omega)]^2 \, .  
\label{EqC6}
\end{equation}
The  coefficient $C_3$
of the interaction between an atom and a perfectly conducting wall
is
\begin{equation}
C_3 = \frac{1}{4 \pi} \int_0^\infty d\omega \; \alpha( i \omega) \, ,  
\label{EqC3}
\end{equation}
or alternatively
\begin{equation}
C_3 = \textstyle{\frac{1}{12}}
     \, \langle v| {\mathbf R} \cdot {\mathbf R} | v \rangle \, .
\label{EqR2}
\end{equation}
Using the latter relation, 
we have previously~\cite{Derevianko98}  determined the values of $C_3$ 
coefficients 
for alkali-metal atoms using  many-body methods.

The dipole operator ${\mathbf R}$, being a one-particle operator,
can have non-vanishing matrix elements for intermediate states
represented by two types of Slater determinant. Firstly, 
the valence electron $v$ can be promoted to some other valence state $w$. 
Secondly,
one of the core orbitals $a$ can be excited to a virtual state $m$, leaving
the valence state $v$ unchanged. In the language of  second-quantization,
the first type of states is represented by  $a_{w}^{\dagger }|0_{c}\rangle$
 and the second type by 
$a_{m}^{\dagger}a_{a}a_{v}^{\dagger }|0_{c}\rangle$, where  
$|0_{c}\rangle$ describes the core. 
These states will be referred to as 
``valence''  and ``autoionizing'' states, respectively.

In accordance with such a classification, 
we break the total polarizability $\alpha$ into
three parts: the polarizability
due to valence states $\alpha_v$, 
the core polarizability $\alpha_c$,
and the valence-core coupling term  $\alpha_{cv}$, with 
\[
\alpha = \alpha_v + \alpha_c + \alpha_{cv} .
\]
The last two terms arise from the summation over autoionizing states.  
In evaluating the core polarizability we permit excitations into
all possible states outside the core. The term $\alpha_{cv}$ is a counter
term accounting for the consequent violation of the Pauli principle.

Various states contribute at drastically different levels to the
dynamic polarizability.  For example, 96\%  of the static polarizability of Cs
is determined  by the two intermediate valence states
$6P_{1/2}$ and $6P_{3/2}$, other valence states contribute less than 1\%.  
The core polarizability accounts for
approximately 4\% of the total value and the contribution of the
core-valence coupling term is about $-0.1$\%.  The relative sizes  of 
contributions to the static polarizabilities for the other
alkali-metal atoms are similar. The dynamic polarizability $\alpha(i \omega)$, 
given in Eq.~(\ref{EqPol}),  behaves as
\[
\alpha (i\omega ) \sim \sum_{k}
  f_{vk}/\omega^{2}= N/{\omega ^{2}} \, ,
\]
at large value of $\omega$, 
where we have used the nonrelativistic oscillator strength sum rule $S\left
( 0\right) =  \sum f_{vk}=N$.
Because the ratio $\alpha_c/\alpha_v$ nonrelativistically is 
close to $N-1$ we 
expect the core polarizability to give the major contribution at large
$\omega$. 
Therefore, the core polarizability becomes
increasingly important for heavier atoms.

Based on the above argument, we use several many-body techniques of
varying accuracy to calculate  the different contributions to the total
polarizability.  In particular, we employed the relativistic
single-double (SD) all-order method to obtain the leading contribution
from valence states~\cite{AllOrder}.  The core polarizability is
obtained from the relativistic random-phase
approximation (RRPA)~\cite{RRPA-pol}. The core-valence coupling term
and the non-leading contribution from valence states is estimated in
the Dirac-Hartree-Fock approximation by a direct summation over basis
set functions~\cite{B-splines}.

The relativistic single-double (SD) all-order method has been
previously used to obtain high-precision atomic properties for the
first few excited states in alkali-atom systems~\cite{AllOrder}.  The
results of theoretical SD matrix elements and comparison with
experimental data are presented elsewhere~\cite{Safronova99}.
Generally, the electric-dipole matrix elements for principal
transitions agree with precise experimental data to better than 0.5\%
for all alkali-metal atoms; the calculations being more accurate for
lighter elements.  In the present work, for Na, K, Rb, and Cs, we have
used SD matrix elements for the first six lowest $P_{1/2}$ and
$P_{3/2}$ levels.
For Fr, we have used SD matrix elements for a principal transition and
matrix elements calculated with the third-order many-body perturbation
theory (MBPT), described in~\cite{thirdorder}, for four other lowest
$P_{1/2}$ and $P_{3/2}$ states.
Unless noted otherwise, we have used experimental values of energy
levels from Ref.~\cite{En_tables} and from the compilation of Dzuba
{\em et al.}~\cite{Dzuba_Fr95} for Fr.

The relativistic random-phase approximation (RRPA) was used previously
to obtain static core polarizabilities for all alkali-metal atoms
except Fr in Ref.~\cite{RRPA-pol}. In the present calculations we
reformulated the original differential equation method used
in~\cite{RRPA-pol} in terms of basis sets~\cite{B-splines}, in a
manner similar to~\cite{WRJ88}. We reproduce the results of
Ref.~\cite{RRPA-pol} and, in addition, obtain a value of 20.41
a.u. for the static dipole polarizability of the Fr$^{+}$ ion.  Zhou
and Norcross~\cite{Zhou89} find $\alpha_c(0)$ = 15.644(5) for the
polarizability of Cs$^{+}$, by fitting Rydberg states energies to a
model potential for Cs, while the present RRPA calculations yield the
value $\alpha_c(0)$=15.81.  Based on this comparison, we expect the
RRPA method to give at least a few per cent accuracy in the
calculation of $\alpha_c(iw)$.

To demonstrate the sensitivity of our results to errors in the core
polarizability, we present the ratios of values calculated omitting
$\alpha_c$ to the total values of $\alpha(0)$, $C_3$, and $C_6$ in
Table~\ref{TabCoreContrib}.  We see that while $\alpha(0)$ is affected
at the level of a few per cent, the core contribution to $C_6$ becomes
increasingly important for heavier systems. $\alpha_c(iw)$ contributes  
 2\% to $C_6$ for Na and  23\% for Fr.  The atom-wall interaction
constant $C_3$, obtained with Eq.~(\ref{EqC3}), is the most sensitive
to the core contribution.  Indeed, while $\alpha_c$ contributes 16\%
of $C_3$ for Na, it accounts for the half of the total value of $C_3$
for Fr.

The tabulation of our results for static dipole polarizabilities,
atom-wall interaction constants $C_3$, and $C_6$ dispersion
coefficients is presented in
Tables~\ref{TabPol}--\ref{TabDisp}. In Method I we use 
high-precision experimental values~\cite{ExpData} for dipole matrix
elements of the principal transition. We used a weighted average of
experimental data if there were several measurements for a particular
transition.  In Method II we use  the theoretical SD matrix
elements for the principal transition. We recommend using the values
obtained with Method I for $\alpha(0)$ and $C_6$, since the accuracy
of experimental data for the principal transitions is better than that
of SD predictions.

In Table~\ref{TabPol} we compare our calculations with experimental data 
for static polarizabilities.  We find perfect agreement
with a high-precision value for Na obtained in recent
atom-interferometry experiments~\cite{Ekstrom96}. The 
experimental data for static polarizabilities of K, Rb, and Cs are
known with the accuracy of about 2\%~\cite{Molof74,Hall74}.  While we
agree with those experimental values, we believe that our theoretical 
approach gives more accurate results, mainly due to the overwhelming
contribution of the principal transition to the sum over intermediate
states. The electric-dipole matrix elements for principal transitions
are known typically at the level 0.1\% accuracy for all alkalis.  The
theoretical error is estimated from the experimental accuracy of matrix
elements~\cite{ExpData}, from an estimated 5\%  error for the
core polarizabilities, and 10\% error for the remaining 
contributions to $\alpha(0)$.

A sensitive test of the quality of the present dynamic
polarizability functions is obtained
by calculating $C_3$ coefficients in two different ways: 
{\em i\/})  by
direct integration of 
$\alpha(i \omega)$ using Eq.~(\ref{EqC3}) and
{\em ii\/})  by calculating
the diagonal 
expectation value of ${\mathbf R}^2$ in Eq.~(\ref{EqR2}). In the present
work we extend calculations of the
expectation value of ${\mathbf R}^2$~\cite{Derevianko98}
in the SD formalism to obtain $C_3$ values for
Rb, Cs, and Fr.  In the Table~\ref{TabWall}, we compare the SD
values for $C_3$ with those obtained in~\cite{Derevianko98} using
MBPT. The difference of 7\% for Cs and 10\% for Fr between SD and
MBPT values is not surprising, since the MBPT~\cite{thirdorder}
underestimates the line-strength of principal transitions by a few per
cent for Cs and Fr.  
To make a consistent comparison between 
the $C_3$ values  obtained by integrating $\alpha(i
\omega)$ and by calculating the expectation value,
 we have used SD energies and matrix elements in  Method II
calculations  in Table~\ref{TabWall}.
These $C_3$ values   agree to about
0.6\% for Na, 1\% for K and Rb, 2.5\% for Cs, and 3.4\% for Fr.  
At present, it appears no experimental data are available for comparison.
We assume that most of the error is due to the RRPA method used
to calculate the core polarizability.
Therefore,  the error estimates in $C_6$ are 
based on the accuracy of experimental 
matrix elements for the principal transition~\cite{ExpData}, and 
by scaling the error of core contribution from $C_3$ to $C_6$,
using Table~\ref{TabCoreContrib}.

The comparison of $C_6$ coefficients with other calculations is 
presented in the Table~\ref{TabDisp}.  
For Na the results are in good agreement with 
a semi-empirical determination~\cite{Kharchenko97}.
The integration over
$\alpha(i\omega)$ 
as in Eq.~(\ref{EqC6}) has been most recently used by Marinescu,
Sadeghpour, and Dalgarno~\cite{Marinescu94} and by Patil and
Tang~\cite{Patil97}. In contrast to the  present {\em ab initio}
calculations, both works employed model potentials.  In addition,
Ref.~\cite{Marinescu94} used corrections to multipole operators to
account for core polarization effects with parameters  chosen to
reproduce the experimental values of static polarizabilities, which
for K, Rb, and Cs atoms are known from
experimental measurements with an accuracy of approximately 2\%.  The
major contribution in the integration of Eq.~(\ref{EqC6})
 arises from the region of
$\omega =0$ and the integrand is quadratic in $\alpha(i\omega)$. 
Therefore, until more accurate experimental values for
static polarizabilities are available, the
predictions~\cite{Marinescu94} of $C_6$ for K, Rb, and Cs have an
inherent (experimental) accuracy of about 4\%.  Theoretical
uncertainty of the method used in Ref.~\cite{Marinescu94} is
determined, among other factors, by the omitted contribution
from core polarizability as
discussed in Refs.~\cite{Marinescu94,Marinescu94C3}.
Patil and Tang~\cite{Patil97} used model-potential calculations with 
analytical representations of wave functions and with experimental
energies. They used a direct summation method in Eq.~(\ref{EqPol}). The
contribution from the core polarizability was not included  as
can be seen from Eq.~(3.4) of Ref.~\cite{Patil97}.  In fact, this
formula in the limit of large $\omega$ results
in $\alpha (i\omega )\rightarrow 1/\omega^2$ instead of the correct 
limit 
$\alpha (i\omega )\rightarrow N/\omega^2$, which follows
from the oscillator strength sum rule.  Therefore, the
model-potential calculations generally underestimate the $C_6$
coefficients.  Indeed, from the comparison in Table~\ref{TabDisp}, one can
see that the $C_6$ values from Ref.~\cite{Marinescu94} and
Ref.~\cite{Patil97} are systematically lower than our values.

Maeder and Kutzellnigg~\cite{Maeder79} used a method alternative to 
the integral Eq.~(\ref{EqC6}) to calculate
dispersion coefficients by minimizing a
Hylleraas functional providing a lower bound.
However, their prediction depended on the quality of the
solution of the Schr\"{o}dinger equation for the ground state. For
alkali-metal atoms, model potentials were used to account for
correlations.  The predicted static polarizabilities are several
per cent higher than experimental values, and are not within the
experimental error limits.  However, for $C_6$ coefficients we generally
find good agreement with the values of Maeder and Kutzellnigg~\cite{Maeder79}.

Recently Marinescu {\em et al.}~\cite{Marinescu98} presented
calculations of dispersion coefficients of different molecular
symmetries for Fr, using a model potential method similar to
Ref.~\cite{Marinescu94}. As shown in Table IV our result for Fr is
significantly larger than the result of Ref.~\cite{Marinescu98}. 
We believe this may be because the method of Ref.~\cite{Marinescu98}
does not completely take into account the contribution
of the core polarizability, which accounts for 23\% of  $C_6$ for Fr.

Elastic scattering experiments and photoassociation spectroscopy have
sensitively constrained  the  possible values of
$C_6$ for Na and Rb.
Van Abeelen and VerHaar~\cite{VanAbeelen99}
reviewed spectroscopic and cold-collision data
for Na, including data from recent observations
of magnetic field induced Feshbach resonances~\cite{Inouye98}.
They considered values for Na of $1539<C_6<1583$ 
and concluded that $C_6=1539$ gave the best consistency
between data sets. Our result for Na using Method~I is
in particularly good agreement with this value.
Photoassociation experiments~\cite{Boesten97} for Rb limits the $C_6$
coefficient to a range 4400-4900 a.u. and even
more recently~\cite{Roberts98} a study of a Feshbach resonance
in elastic collisions of ${}^{85}$Rb 
concluded $C_6=4700(50)$.  Our value $C_6=4691(23)$ is in excellent
agreement with this experiment. 
For Cs, knowledge of the value of $C_6$ is critical for predictions of
the sign of the elastic scattering length~\cite{Grubellier98}, though
it has been demonstrated the
resulting cross sections are not particularly sensitive
to the value of $C_6$~\cite{Leo98}.  For Fr, the paucity of other
dimer data constrains quantitative theoretical
collisional studies for the near future.  As photoassociation
experiments move beyond the alkali-metal atoms to other atoms with
many electrons such as Sr~\cite{Dineen99} and Cr~\cite{Bradley98}, it
will be important to have reliable {\em ab initio\/} methods for
calculation of atomic properties. The approaches presented here could,
in principle, be applied to Sr and perhaps with some significant effort
to Cr.

AD would like to thank 
H.\ R.\ Sadeghpour, B.\ D.\ Esry, F.\ Masnou-Seeuws, 
and D.\ J.\ Heinzen for useful discussions.
The work of AD, WRJ, and MSS was supported in part by 
NSF Grant No.\ PHY 95-13179 and that of JFB by NSF Grant No. PHY 97-24713. 
The Institute for Theoretical
Atomic and Molecular Physics is supported by a grant from the NSF
to the Smithsonian Institution and Harvard University.

\begin{table}
\caption{Demonstration of the  relative importance of
the  contribution
of autoionizing states with increasing 
number of electrons $N$, where $\alpha^{v}$,  $C_3^{v}$,  
and $C_6^{v}$ represent values calculated disregarding
autoionizing states.
 \label{TabCoreContrib} }
\begin{tabular}{llllll}
       & Na &  K & Rb & Cs  &  Fr \\ 
\hline
$\alpha^v(0)/\alpha(0)$
      & 0.99  & 0.98  & 0.97 &  0.96  & 0.94 \\
$C_3^v/C_3$
      & 0.84  & 0.73  & 0.65 &  0.59  & 0.50 \\
$C_6^v/C_6$ 
      & 0.98  & 0.93  & 0.89  & 0.85  & 0.77  \\
\end{tabular}
\end{table}

\begin{table}
\caption{  Comparison of 
static  dipole polarizabilities $\alpha(0)$ 
for alkali-metal atoms in atomic units. Method I designates 
the use of high-accuracy experimental data for
electric-dipole matrix elements for principal transition.
Method II designates the use of all-order SD values instead.  
 \label{TabPol} }
\begin{tabular}{lllllll}
       & Na &  K & Rb & Cs & Fr  \\ 
\hline
Method I\tablenotemark[1] 
      &  162.6(3)   & 290.2(8)    & 318.6(6) & 399.9(1.9)  &	317.8(2.4) \\
Method II 
      & 163.0        & 289.1     & 316.4    & 401.5       &    315.1	\\
Expt.~\cite{Ekstrom96}  
      & 162.7(8) \\
Expt.~\cite{Molof74,Hall74}\tablenotemark[2]     
      &         & 293.6(6.1) & 319.9(6.1) &  403.6(8.1)  &     
\end{tabular}
\tablenotetext[1]{ Values recommended from the present work.}
\tablenotetext[2]{ Weighted average of experimental data from 
Refs.~\cite{Molof74,Hall74}.
}
\end{table}

\begin{table}
\caption{  Comparison of 
atom-wall interaction
constants $C_3$
for alkali-metal atoms in atomic units. Method I designates 
the use of high-accuracy experimental data for
electric-dipole matrix elements and energies for principal transition.
Method II designates the use of all-order SD values instead.  
 \label{TabWall} }
\begin{tabular}{lllllll}
       & Na &  K & Rb & Cs & Fr  \\ 
\hline
Method I, Eq.~(\protect\ref{EqC3})
      &  1.871  & 2.896 & 3.426 & 4.269    &  4.437      \\
Method II, Eq.~(\protect\ref{EqC3}) 
     & 1.875   & 2.887 &  3.410 & 4.247    &  4.427   \\
$\frac{1}{12} \langle R^2 \rangle$, SD\tablenotemark[1]\tablenotemark[2],
Eq.~(\protect\ref{EqR2})
     & 1.8858 & 2.860  &  3.362 & 4.143    &  4.281   \\
$\frac{1}{12} \langle R^2 \rangle$, MBPT~\cite{Derevianko98},
Eq.~(\protect\ref{EqR2})
     & 1.8895 & 2.838  &  3.281 & 3.863    &  3.870     
\end{tabular}
\tablenotetext[1]{ Values recommended from the present work.}
\tablenotetext[2]{ The values for Na and K are from 
Ref.~\cite{Derevianko98}, and those for Rb, Cs, and Fr
are the present calculations.
}
\end{table}

\begin{table}
\caption{ Tabulation and comparison of 
$C_6$ dispersion coefficients
for alkali-metal atoms in atomic units. Method I designates 
the use of high-accuracy experimental data for
electric-dipole matrix elements for principal transition.
Method II designates the use of all-order SD values instead.  
 \label{TabDisp} }
\begin{tabular}{lllllll}
       & Na &  K & Rb & Cs & Fr  \\ 
\hline
Method I\tablenotemark[1]  
      & 1556(4)  & 3897(15) & 4691(23)  & 6851(74)   &   5256(89)      \\
Method II 
      & 1564    & 3867     & 4628  & 6899    & 5174   \\
Ref.~\cite{Kharchenko97}\tablenotemark[2]
     & 1561   &   &    &     &    \\
Ref.~\cite{Maeder79} 
     & 1540   & 3945  & 4768 & 6855   &        \\
Ref.~\cite{Marinescu94} 
     & 1539\tablenotemark[3]   & 3813  & 4426 & 6331  &       \\
Ref.~\cite{Patil97}  
     & 1500  & 3796  & 4531   & 6652  &        \\
Ref.~\cite{Marinescu98}  
                       &    &   &   &   &  3934\tablenotemark[4] \\
Expt.~\cite{Boesten97} 
     &          &       & 4400-4900 &       &  \\ 
Expt.~\cite{Roberts98} 
     &          &       & 4700(50) &       &  
\end{tabular}
\tablenotetext[1]{ Values recommended from the present work.}
\tablenotetext[2]{ Semiempirical method.}
\tablenotetext[3]{ For Na the value from Ref.~\protect\cite{Marinescu94}
is 1472,
obtained using the data from Ref.~\protect\cite{Molof74}. Using the same
method, but with data from Ref.~\protect\cite{Ekstrom96}, the resulting
value is 1539~\protect\cite{Marinescu94C3}.}
\tablenotetext[4]{ Value for $^3\!1_u$ molecular symmetry.
Values for other symmetries are $C_6(^1\!0_g^+)$ = 3929, 
and $C_6(^1\!0_u^-)$ = 3947.
}
\end{table}


\begin{references}
\bibitem{Yan98} Results for Li 
already have been presented 
using a precise
nonrelativistic {\em ab initio\/} approach, see Z.-C. Yan,
A. Dalgarno, and J. F. Babb, Phys. Rev. A {\bf 55}, 2882 (1997).

\bibitem{Derevianko98} A. Derevianko, W. R. Johnson, and S.  Fritzshe,
 Phys.\ Rev.\ A{\bf 57}, 2629 (1998).

\bibitem{AllOrder} S.\ A.\ Blundell, 
W.\ R.\ Johnson,  Z.\ W.\ Liu and J.\ Sapirstein,
Phys.\ Rev.\ A {\bf 40}, 2233 (1989);
M. S. Safronova, A. Derevianko, W. R. Johnson,
Phys.\ Rev.\ A{\bf 58}, 1016 (1998).

\bibitem{RRPA-pol} D. Kolb, W.\ R.\ Johnson, and P. Shorer,
 Phys.\ Rev.\ A{\bf 26}, 19 (1982);
W.\ R.\ Johnson, D. Kolb, and K.-N.\ Huang, 
At. Data Nucl. Data Tables {\bf 28}, 333 (1983).

\bibitem{B-splines} W.\ R.\ Johnson, S.\ A.\ Blundell, and J. Sapirstein,
Phys.\ Rev.\ A{\bf 37}, 307 (1988).

\bibitem{Safronova99} M.\ S.\ Safronova, A.\ Derevianko, 
and W.\ R.\ Johnson (unpublished).

\bibitem{thirdorder} W. R. Johnson, Z. W. Liu, and J. Sapirstein, 
At. Data Nucl. Data Tables {\bf 64}, 280 (1996).

\bibitem{En_tables} C. E. Moore, Atomic Energy Levels, NBS Ref. Data Series,
1971, Vol.I-III.
\bibitem{Dzuba_Fr95} V.\ A.\ Dzuba, V.\ V.\ Flambaum, and O.\ P.\ Sushkov,
Phys.\ Rev.\ A{\bf 51}, 3454 (1995).

\bibitem{WRJ88} W.R. Johnson, Adv.  At. Mol. Phys., 
D. Bates, B. Bederson, eds. (Academic, Boston, 1988) {\bf 25}, 375.

\bibitem{Zhou89} H. L. Zhou and D. W. Norcross, 
Phys.\ Rev.\ A{\bf 40}, 5048 (1989).
\bibitem{ExpData} 
U. Volz and H. Schmoranzer, Phys. Scr. T {\bf 65}, 48 (1996); 
K.\ M.\ Jones, P.\ S.\ Julienne, P.\ D.\ Lett, W.\ D.\ Phillips, E.\ Tiesinga, 
and C.J. Williams, Europhys. Lett. {\bf 35}, 85 (1996);
H.\ Wang, P.\ L.\ Gould,  W.\ C.\ Stwalley, J. Chem. Phys. 
{\bf 106} 7899 (1997);
R.\ S.\ Freeland {\em et al.} (unpublished);
J.\ E.\ Simsarian, L.\ A.\ Orozco, G.\ D.\ Sprouse, W.\ Z.\ Zhao, Phys. Rev. A 
{\bf 57} 2448 (1998);
L.\ Young, W.\ T.\ Hill III, S.\ J.\ Sibener, S.\ D.\ Price, C.\ E.\ Tanner,
C.\ E.\ Wieman, and S.\ R.\ Leone, Phys. Rev. A {\bf 50}, 2174 (1994).
R.J. Rafac, C. E. Tanner, A. E. Livingston, and H. G. Berry 
(submitted to PRA, 1998); 
R. J. Rafac, C. E. Tanner, A.E. Livingston, K. W. Kukla, H. G. Berry, and C. A. Kurtz, 
Phys. Rev. A {\bf 50}, R1976 (1994).

\bibitem{Ekstrom96} C. R. Ekstrom, J. Schmiedmeyer, M. S. Chapman,
T. D. Hammond, and D. E. Pritchard, Phys. Rev. A{\bf 51}, 3883 (1996).
\bibitem{Molof74} R. W. Molof, H. J. Schwartz, T. M. Miller, and Bederson,
Phys.\ Rev.\ A{\bf 10}, 1131 (1974).
\bibitem{Hall74} W. D. Hall, J. C. Zorn, 
      Phys.\ Rev.\ A {\bf 10}, 1141  (1974).

\bibitem{Kharchenko97} P. Kharchenko, J. F. Babb, and A. Dalgarno, Phys.\ Rev.\
A{\bf 55}, 3566 (1997).
\bibitem{Marinescu94} M.\ Marinescu, H.\ R.\ Sadeghpour, and A.\ Dalgarno,
Phys.\ Rev.\ A{\bf 49}, 982 (1994).
\bibitem{Patil97} S. H. Patil, K. T. Tang, J. Chem. Phys., {\bf 106}, 
2298 (1997).
\bibitem{Marinescu94C3}  M.\ Marinescu, J.\ F.\ Babb, and A.\ Dalgarno,
Phys.\ Rev.\ A{\bf 50}, 3096 (1994).

\bibitem{Maeder79} F. Maeder and W. Kutzellnigg, 
Chem. Phys.{\bf 42}, 195 (1979). 


\bibitem{Marinescu98} M.\ Marinescu, D. Vrinceanu, and H.\ R.\ Sadeghpour,
Phys.\ Rev.\ A{\bf 58}, R4259 (1998).

\bibitem{VanAbeelen99} F. A. VanAbeelen and B. J. Verhaar, 
Phys. Rev. A{\bf 59}, in press (1999).
\bibitem{Inouye98} S. Inouye, M. R. Andrews,
J. Stenger, H.-J. Miesner, D. M. Stamper-Kurn,
and W. Ketterle, Nature {\bf 392}, 151 (1998).

\bibitem{Boesten97} H. M. J. M. Boesten, C. C. Tsai, J. R. Gardner,
D. J. Heinzen, and B. J. Verhaar, Phys. Rev. A{\bf 55}, 636 (1997).  
\bibitem{Roberts98} J. L. Roberts, N. R. Claussen, J. P. Burke, Jr.,
C. H. Greene, E. A. Cornell, and C. E. Wieman,
Phys. Rev. Lett {\bf 81}, 5109 (1998).  
\bibitem{Grubellier98} A. Grubellier, O. Dulieu, 
F. Masnou-Seeuws, M. Elbs, H. Knockel, and E. Tiemann,
(subm. to Eur. J. Phys., 1998).
\bibitem{Leo98} P. J. Leo, E. Tiesinga, P. S. Julienne,
D. K. Walter, S. Kadlecek, and T. G. Walker, Phys. Rev. Lett. {\bf 81},
1389 (1998).
\bibitem{Dineen99} T. P.
Dinneen, K. R. Vogel, J. L. Hall, and A. Gallagher,
Phys. Rev. A{\bf 59}, in press (1999).
\bibitem{Bradley98} C. C. Bradley, W. R. Anderson, J. J. McClelland,
and R. J. Celotta, BAPS {\bf 43}, 1291 (1998).
\end{references}
\end{document}